\documentclass[%
reprint,
 amsmath,amssymb,
 prl,
]{revtex4-2}
\usepackage{graphicx}
\usepackage{dcolumn}
\usepackage{bm}

\begin{document}

\preprint{APS/123-QED}

\title{Ponderomotive electron physics captured in a single-fluid extended magnetohydrodynamics model}

\author{James R. Young}
\author{Pierre-Alexandre Gourdain}%
\affiliation{%
 Department of Physics and Astronomy, University of Rochester, Rochester, New York 14627, USA\\
}%
\affiliation{Laboratory for Laser Energetics, University of Rochester, Rochester, New York 14627, USA\\}

\date{\today}

\begin{abstract}
 The well-known ponderomotive force, arising from the interaction of a plasma with electromagnetic waves, has critical implications across a broad range of fields from laser fusion and astrophysics to laser diagnostics and even pulsed-power experiments. This force pushes electrons, which causes steepening ion density modulations through Coulomb forces.   When used intentionally, density modulations can be used for plasma gratings, which are essential for optical components that operate under extreme conditions. Ponderomotive forces can also lead to laser self-focusing, which complicates laser diagnostics.  It also enhances stimulated Brillouin scattering (SBS) and crossed-beam energy transfer (CBET), both of which reduce the convergence in laser fusion. It even plays a role in the filamentation of fast radio bursts in the relativistic winds of magnetars.  Since the ponderomotive force primarily effects electron dynamics, multi-fluid or particle codes are required to study its effects. This paper demonstrates that by including electron effects in the generalized Ohm's law, a 1-fluid, 2-energy extended magnetohydrodynamics (XMHD) model can capture ponderomotive effects accurately. We introduce phasor theory to describe the ponderomotive force in this framework.  
\end{abstract}


\maketitle

\section{Introduction}
The ponderomotive force is a non-linear force caused by the beating of collective plasma and electromagnetic waves.  It has important consequences for fusion energy, laser diagnostics, astrophysics, and laser-triggered x-pinches\cite{young_using_2021}.  Ponderomotive force leads to modulations in electron and ion density that eventually increase \cite{estabrook_two-dimensional_1975, smith_particle--cell_2019, dragila_laser_1978}.  It has also been presented as a source of magnetization in plasma\cite{gradov_magnetic-field_1983, shukla_magnetization_2009}.  The sharp nature of these modulations can even be used to form plasma gratings\cite{lehmann_plasma_2019, sheng_plasma_2003, plaja_analytical_1997, edwards_plasma_2022, peng_nonlinear_2019} when optical components must operate in extreme conditions.  Petawatt lasers \cite{danson_petawatt_2015} require diffraction optics, which cannot withstand energy densities larger than $\approx 1\mathrm{\frac{J}{cm^2}}$\cite{chambonneau_multi-wavelength_2018}.  Even if another mechanism is used to form the plasma grating, ponderomotive effects will likely influence the optics.  Density modulations have also been introduced as a consequence of ponderomotive laser self-focusing\cite{kaw_filamentation_1973, del_pizzo_self-focussing_1979} and even applied to the filamentation of fast radio bursts in the relativistic winds of magnetars\cite{sobacchi_saturation_2023, ghosh_nonlinear_2022}.  Additionally, parametric instability\cite{kruer_physics_2003} (stimulated Brillouin scattering) combined with ponderomotively driven plasma response has long been a major concern for laser fusion\cite{bezzerides_convective_1996, eliseev_stimulated_1995, myatt_multiple-beam_2014} and can even contribute to crossed beam energy transfer (CBET)\cite{huller_crossed_2020, ruocco_modelling_2020} which further destabilizes laser fusion.   

However, resolving ponderomotive dynamics is particularly challenging as it involves spatial scales on the order of the laser wavelength $\lambda_L$.  Laser effects are generally included in magnetohydrodynamics (MHD) and hybrid fluid-kinetic models using the paraxial approximation and incorporating ray-tracing \cite{kaiser_laser_2000,tzeferacos_flash_2015,xu_ion_2024}. However, in the presence of ponderomotive effects and turbulent density fluctuations from high-density plasma such as x- and z-pinches \cite{klein_scaling_1993, rocco_turbulence, rocco2022, kroupp_ion_2011, kroupp_turbulent_2018}, this assumption can break down \cite{hyole_validity_2000}. Although MHD cannot generally handle electron physics, it is possible to approximate electron effects on an ion time-scale using the generalized Ohm's law coupled to a relaxation scheme \cite{seyler_relaxation_2011, martin_2010}, to form the extended MHD (XMHD) model on which the code PERSEUS is built.  Usually, fast electron motion is neglected as it is not typically relevant to the MHD approximation of a plasma.  However, by removing it along with the associated displacement current ($\mathrm{\frac{\partial E}{\partial t}}$) from Maxwell-Ampere's law, any hope of retaining electromagnetic waves in a plasma is broken.  

We will show in this paper how electron effects averaged over the ion-timescale using the GOL can be retained in a single fluid extended MHD in the presence of an electromagnetic (EM) wave.  Including those effects recovers some of the ponderomotive effects discussed above.  For example, stimulated Brillouin scattering (SBS) turns an EM wave into another EM wave along with an ion acoustic wave (IAW), both of which are supported by the XMHD model.  Previous work \cite{young_impact_2024} has demonstrated consistency with collisional absorption of EM laser energy, as well as 2D effects \cite{young_using_2021} like filamentation and self-focusing.  The code speedup was nearly 10$\times$ compared to equivalent Particle-in-cell (PIC) models in that case.

We first introduce the analytic theory for ponderomotive dynamics on a single fluid XMHD ion-timescale.  Then we develop and analyze 1-D simulations demonstrating a match of theory and computation.      

\section{Redimensionalization of XMHD equations}
In the context of laser-plasma interactions, we take the electromagnetic wave speed $c$ as our characteristic speed $v_0=c$.  Using Faraday's law, we let $E_0=cB_0$. In addition,
the characteristic thermal pressure is $p_0=\rho_0c^2$, the characteristic resistivity is $\eta_0=t_0/\varepsilon_0$,
the characteristic magnetic field is $B_0=\sqrt{\rho_0/\varepsilon_0}$, and the characteristic current density is $j_0=\sqrt{\rho_0/\mu_0}/t_0$.  Additionally, $\omega_{ce0}=\sqrt{eB_0/m_e}$ is the characteristic cyclotron frequency.  For now, the characteristic time, $t_0$, is left undefined, but later the laser period will define it. This all leads to XMHD equations in dimensionless form,  
\begin{equation}\label{eq:mass_eq}
\mathrm{\frac{\partial \rho}{\partial t}+\nabla(\rho \textbf{u})=0}
\end{equation}
\vspace{-2em}
\begin{equation}\label{eq:momentum_eq}
\mathrm{\quad\frac{\partial \rho \textbf{u}}{\partial t}+\nabla \cdot(\rho \textbf{u}\textbf{u}+p\textbf{I})=\textbf{j} \times \textbf{B}}
\end{equation}
\vspace{-2em}
\begin{eqnarray}\label{eq:nrg_eq_ion}
\mathrm{\frac{\partial \epsilon_i}{\partial t}+\nabla \cdot\left(\left[\epsilon_i+p_{i}\right] \textbf{u}\right)=}\nonumber\\
\mathrm{Zn_i\textbf{E}\cdot\textbf{u}+3\frac{1}{m}\nu_en_e(T_e-T_i)-\frac{1}{\omega_{ce0}t_0}\nu_e\mathbf{j} \cdot \mathbf{u}}\label{eq:consengion}
\end{eqnarray}
\vspace{-2em}
\begin{eqnarray}\label{eq:nrg_eq_e}
\mathrm{\frac{\partial \epsilon_e}{\partial t}+\vec{\nabla} \cdot[(\epsilon_e+p_{e})\mathbf{u_e}]= }\nonumber\\
-\mathrm{n_e\textbf{E}\cdot\mathbf{u_e}-3\frac{1}{m}\nu_en_e(T_e-T_i)+\frac{1}{\omega_{ce0}t_0}\nu_e\mathbf{j} \cdot \mathbf{u}} \label{eq:consenge}
\end{eqnarray}

\noindent which is used together with Maxwell's equations and a quasi-neutral plasma assumption.

Using the previously defined characteristic scales we can also write the dimensionless generalized Ohm's law (GOL), 
\begin{eqnarray}\label{eq:dimensionless_gol}
    \mathrm{\mathbf{E}+\mathbf{u} \times \mathbf{B}=\eta\textbf{j}+\frac{1}{\omega_{pe}t_0}\sqrt{\frac{m}{n_e}}(\textbf{j}\times\textbf{B}-\nabla p_e)}\nonumber\\ 
    + \mathrm{\left(\frac{1}{\omega_{pe}t_0}\right)^2\left(\frac{\partial \mathbf{j}}{\partial t}+\nabla \cdot (\mathbf{uj+ju})\right)}\\
    -\mathrm{\left(\frac{1}{\omega_{pe}t_0}\right)^3\sqrt{mn_e}\,\nabla\cdot\left(\frac{1}{n_e}\mathbf{jj}\right)}.\nonumber 
\end{eqnarray}
Here $n_e$ is the electron plasma density, $m$ is the ratio of the ion mass to the electron mass, $\rho$ is the mass density, \textbf{v} is the flow speed, $p$ is the thermal pressure, \textbf{j} is the current density, \textbf{B} is the magnetic field, $\nu$ is the electron-ion collision rate, $\eta$ is the electric resistivity and $\epsilon$ is the energy density, all dimensionless. The redimensionalization of the GOL made the \textit{dimensional} electron plasma frequency $\mathrm{\omega_{pe}=\sqrt{N_ee^2/(m_e\varepsilon_0)}}$ appear in front of all terms connected to electron physics. Here, $N_e$ is electron plasma density, e is the elementary charge, and $m_e$ is the electron mass, all \textit{dimensional}.

\section {Description of the Ponderomotive force using phasors}

Since this work aims to demonstrate the existence of the ponderomotive force ($\mathbf{F}_\mathrm{p}$) within the XMHD model, we now define its dimensionless form.  

\begin{equation}
    \mathrm{\mathbf{F}_\mathrm{{p}}=\frac{1}{4}\left(\omega_{pe}t_0\right)^{2}\nabla(\left\Vert \mathbf{E}\right\Vert^{2})}
    \label{eq:Fpond}
\end{equation}

The generalized Ohm's law is the natural starting point for ponderomotive effects as it is a simplification of the electron momentum conservation equation.  

The dyadic current terms in Eq. (\ref{eq:dimensionless_gol}) can be simplified as a consequence of quasi-neutrality ($\mathbf{\nabla \cdot j} = 0$) and several vector identities to yield the following relationship.

\begin{flalign}
    \nabla\cdot(\mathbf{jj}) = \frac{1}{2}\nabla(\left\Vert \mathbf{j}\right\Vert^{2})-\mathbf{j\times\left(\nabla\times j\right)} \label{eq:dyadic_simp}
\end{flalign}

The RHS of Eq. (\ref{eq:dyadic_simp}) has two components - the first, $\frac{1}{2}\nabla(\mathbf{j\cdot j})$, looks functionally similar to the RHS of Eq. (\ref{eq:Fpond}) and the second, $\mathbf{j\times\left(\nabla\times j\right)}$, is similar to the Hall term ($\mathbf{j}\times\mathbf{B}$) in Eq. (\ref{eq:dimensionless_gol}) when the displacement currents are not important.  This term represents the electric field that can develop when ions and electrons decouple in a magnetic field.  We now introduce a phasor representation of this equation to expand on this similarity.  The use of phasors requires assuming current density, electric and magnetic fields vary sinusoidally, $\mathrm{\mathbf{\underline{j} = \tilde{j}(\mathbf{x})} e^{i\omega_Lt}}$, $\mathrm{\mathbf{\underline{E} = \tilde{E}(\mathbf{x})} e^{i\omega_L t}}$, and $\mathrm{\mathbf{\underline{B} = \tilde{B}(\mathbf{x})} e^{i\omega_L t}}$, and that the system has reached steady-state driven by a laser with $\omega_L$ as the dimensionless laser frequency. Additionally, the external laser is also described similarly with $\mathrm{\mathbf{\underline{E}_L = \tilde{E}_L(\mathbf{x})} e^{i\omega_L t}}$, and $\mathrm{\mathbf{\underline{B}_L = \tilde{B}_L(\mathbf{x})} e^{i\omega_L t}}$.  We denote the \textit{complex conjugate} with *.  These assumptions are used both for demonstration purposes and for early-time simulations with no spatial gradients or nonuniformities.

With these definitions, the phasor form of the first part of Eq. (\ref{eq:dyadic_simp}) is shown below. Note that the real component of each phasor equation is implied when we return to physical variables.
                        
\begin{equation}
    \frac{1}{2}\nabla(\mathbf{\Re(\underline{j})\cdot\Re(\underline{j})}) = \frac{1}{4}\nabla(\left\Vert \mathbf{\tilde{j}}\right\Vert^{2}+\frac{1}{2}\mathbf{\tilde{j}\cdot \tilde{j}^*}\mathrm{e^{\pm2i\omega_L t}}) 
    \label{eq:dyadic_phasor0}
\end{equation}

To simplify the second part of Eq. (\ref{eq:dyadic_simp}), some relationship between $\mathbf{E}$ and $\mathbf{j}$ must be defined.  Fortunately, Eq. (\ref{eq:dimensionless_gol}) can relate their fast and slow components using the assumptions already stated.

\begin{flalign}
    &\mathbf{\tilde{E}}+\mathbf{u\!\times\!\tilde{B}}\!=\!\frac{1}{2}\frac{\mathbf{\tilde{j}\!\times\!\tilde{B}^*}}{t_0\omega_{pe}}\mathrm{\sqrt{\frac{m}{n_e}}}\nonumber &\\ 
    & -\frac{1}{2}\!\mathrm{\left(\frac{1}{t_0\omega_{pe}}\right)^{\!3\!}\!\sqrt{\frac{m}{n_e}}\nabla\!\cdot\!(\mathbf{\tilde{j}\tilde{j}^*})\!-\!\sqrt{\frac{m}{n_e}}\frac{\nabla p_e}{t_0\omega_{pe}}}\nonumber &\\ 
    & +\!\mathrm{\!e^{\!-i\omega_Lt\!}\!\left(\!\frac{1}{\!t_0\omega_{pe}}\!\right)^{2}\!\nabla\!\cdot\!(\mathbf{\!u\tilde{j}\!+\!\tilde{j}u\!})\!} \nonumber &\\
    & +\!\mathrm{\!e^{\!-i\omega_Lt\!}\!\left(\!(\mathrm{(\omega_{pe}t_0)^2}\eta\!-\!\mathrm{\omega_L}i)\mathbf{\tilde{j}}\!-\!\mathrm{(\omega_{pe}t_0)^2}\mathbf{\tilde{E}_L} \right) }\nonumber&\\ 
    & +\!\mathrm{e^{\!\pm2i\omega_L t\!}\frac{1}{2}\sqrt{\frac{m}{n_e}}\left(\frac{\mathbf{\tilde{j}\!\times\!\tilde{B}}}{t_0\omega_{pe}}-\frac{\mathbf{\nabla\!\cdot\!(\tilde{j}\tilde{j})}}{(t_0\omega_{pe})^{3}}\right)} \label{eq:gol_phasor}
\end{flalign} 


Note that parts without an $\mathrm{e^{-in\omega_Lt}}$ multiplier represent terms whose average is not necessarily zero in the phasor model. Since all the components oscillating with the same frequency as the laser are contained in the $\mathrm{e^{-i\omega_Lt}}$ term, the same is true for the left and right sides of Eq. (\ref{eq:gol_phasor}) to balance, the following relationship must hold.

\begin{equation}
    (\mathrm{\left(\omega_{pe}t_0\right)^2\eta-\omega_Li})\mathbf{\tilde{j}}=\mathrm{\left(\omega_{pe}t_0\right)^2}\mathbf{\tilde{E}_L}
    \label{eq:jtoE}
\end{equation}

This assumes ions are immobile ($\mathbf{u}\approx0$), which is true early in time and on the timescale of the laser oscillation.  These conditions also imply $\mathrm{1/t_0\approx\omega_L}$, which in turn means $\mathrm{\omega_L\approx1}$.  This relationship, along with Faraday's law, implies $\mathbf{\nabla\times \tilde{j}}$ may now be recast as a function of only $\mathbf{\tilde{B}_{L}}$.  Consequently, when $\mathbf{\tilde{j}\times\left(\nabla\times \tilde{j}\right)}$ is combined with Eq. (\ref{eq:dyadic_phasor0}), then we fully define Eq. (\ref{eq:dyadic_simp}) in the phasor form. 

\begin{flalign}
    \langle\nabla\cdot(\mathbf{\underline{j}\underline{j}})\rangle & = \frac{1}{4}\nabla(\left\Vert \mathbf{j}\right\Vert^{2}) \nonumber \\
    & +\mathrm{\frac{1}{2}\left(\left(\frac{a^2n_e}{a^4n_e^2\eta^2\!+\!1}\right)\!+\!i\left(\frac{a^4n_e^2\eta}{a^4n_e^2\eta^2\!+\!1}\right)\right)\mathbf{\tilde{j}\times\tilde{B}^{*}}} \nonumber \\
    \label{eq:dyadic_phasor1}
\end{flalign}

Here we used $\mathrm{a=\omega_{pe}t_0}$ to visually simplify Eq. (\ref{eq:dyadic_phasor1}).  Additionally, we have taken an average over one laser cycle, where the $\mathrm{<>}$ indicates a variable has been averaged over the wave period.  One final simplification we make is low-resistivity, $\eta<<1$, which makes $\langle\nabla\cdot(\mathbf{\underline{j}\underline{j}})\rangle\approx\frac{1}{4}\nabla(\left\Vert \mathbf{j}\right\Vert^{2})+\frac{a^2n_e}{2}\mathbf{\tilde{j}\times\tilde{B}^{*}}$.  This is justified since lasers rapidly heat colder plasma at early times as mentioned in prior work \cite{young_impact_2024}.  

Although we will not include a detailed exploration of resistivity in this work, it is well-known that electron distribution functions can quickly become non-Gaussian.  However, corrections can be made to resistivity models to account for non-Gaussian absorption effects \cite{matte_non-maxwellian_1988} and, more recently, for ion screening effects \cite{turnbull_inverse_2023} on absorbed collisional energy.   

Substituting this simplified form of the dyadic term into the GOL in Eq. (\ref{eq:gol_phasor}) demonstrates how the ponderomotive term enters the extended MHD model.  

\begin{equation}
    \langle\mathbf{\tilde{E}}+\mathbf{u\!\times\!\tilde{B}}\rangle\!=\!-\frac{1}{4}\!\mathrm{\sqrt{\frac{m}{n_{e}}}}\left(\mathrm{\frac{1}{t_0\omega_{pe}}}\right)^{\!3\!}\nabla\left\Vert \mathbf{j}\right\Vert^{2}\!-\!\mathrm{\sqrt{\frac{m}{n_e}}\frac{\nabla p_e}{t_0\omega_{pe}}}
    \label{eq:GOLPhasorDC}
\end{equation} 

If we again substitute $\mathbf{\tilde{E}}_\mathrm{L}$ for $\mathbf{\tilde{j}}$  using Eq. (\ref{eq:jtoE}) and write this as a momentum, Eq. (\ref{eq:momentum_eq}), then the form of $\mathbf{F}_{\mathrm{p}}$ in Eq. (\ref{eq:Fpond}) is recovered.  

\begin{equation}
    \mathrm{\sqrt{\frac{n_e}{m}}t_0\omega_{pe}}\langle\mathbf{\tilde{E}}+\mathbf{u\!\times\!\tilde{B}}\rangle\!=\!-\mathrm{\frac{1}{4}\!(\omega_{pe}t_0)^2}\nabla\left\Vert \mathbf{E_L}\right\Vert^{2}\!-\!\nabla \mathrm{p_e}
    \label{eq:FpondInPhasor}
\end{equation} 

The ponderomotive force is supported by XMHD through the inclusion of electron inertia and the Hall-term of the generalized Ohm's law.  Furthermore, Eq. (\ref{eq:FpondInPhasor}) shows $\mathrm{\mathbf{F}_{p}}$ can either change $\mathrm{p_e}$ or generate a static electric or magnetic field.  As the model described in Eqns. \ref{eq:mass_eq}-\ref{eq:nrg_eq_e} suggests, there are only two mechanisms the ions can feel $\mathrm{\mathbf{F}_{p}}$, either through $\mathbf{j}\times\mathbf{B}$, through $\nabla\cdot(\mathrm{p_i+p_e})\mathbf{I}$, or through a combination of both.   

The expected ion-fluid velocity for this time-averaged force can be estimated with some simplifying assumptions.  If $\nabla \langle\mathbf{E_L\cdot E_L}\rangle\propto \mathrm{sin(2k_Lx)}$, then the ion velocity for the journey from $\mathrm{x=\lambda_L/8}$ to $\mathrm{3\lambda_L/16}$ can be defined.  It corresponds to the maximum and halfway to the minimum locations of the ponderomotive force.  This is performed by the following integral,

\begin{eqnarray}
\mathrm{\int_{0}^{u_{max}} v \,dv= \int_{\lambda_L/8}^{3\lambda_L/16}\frac{F_{p}}{\rho} \,dx} \nonumber\\
\mathrm{u_{max} = \frac{E_0e\lambda_L}{2^{5/4}\pi m_ec}\sqrt{Z/m}}.
\label{eq:velocity_int}
\end{eqnarray}

In the dimensional Eq. (\ref{eq:velocity_int}), \textit{m} refers to $\mathrm{m_{i}/m_e}$, Z is the ionization state of the plasma and the dimensional mass density ($\rho$) is assumed to be slowly varying in space.  The equation represents the velocity ions would attain if traversing the distance from the highest pondermotive potential to halfway to the lowest.  The integration cannot go to the minimum since actual $\mathrm{u_{max}}$ at $\mathrm{x=\lambda_L/4}$ will be $\approx 0$ as there are equal and opposing forces pushing ions toward this location.  The ion density and pressure should also peak here, which suggests the simple model used in Eq. (\ref{eq:velocity_int}) does not capture all the physics.

\section{Comparison of XMHD and other models}

\begin{figure}[h!]
\centerline{\includegraphics[]{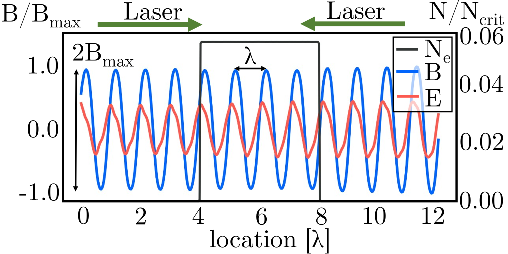}}
\caption{\label{fig:ezby_snapshot} A snapshot of electron density along with the normalized magnetic and electric fields. }
\end{figure}

Since this work has shown a ponderomotive force is present in single fluid XMHD, the model will now be used to generate a simple ponderomotive ion-driven plasma \textit{grating}.  Although previous work \cite{young_impact_2024} has shown a remarkable consistency between this XMHD model and Particle-in-Cell (PIC) models, that analysis was primarily focused on collisional laser absorption.  This work will now expand on collisionless absorption (i.e. absorption not reliant on particle collisions). We use these simulations to compare XMHD with a typical analytical model used to validate PIC models, including electrostatic fields and electron oscillations \cite{smith_particle--cell_2019, sheng_plasma_2003, plaja_analytical_1997}. This section also returns to \textit{dimensional} equations.

 We limit the physics to 1-D, non-relativistic linearly s-polarized electromagnetic fields.  The equations are fully solved in 2D, but the resolution is along the x-direction, with only 9 cells in the y-direction.  Furthermore, the ($12\lambda$) domain was constructed such that a standing wave will form in a low-density plasma.  This simulation includes two identical counter-propagating laser beams propagating along the +/- x-axis towards a low-density ($\mathrm{N_{e}/N_{crit} = 0.05}$) plasma.  The magnetic field components, $B_{y}$, were defined on the vertical boundaries so the Poynting flux was directed to the left and right.  The beam has a uniform spatial profile, $\mathrm{B_{y}=B_{max}\cos{\left(k_Lx+\omega_L t\right)}}$.  Here, $\mathrm{B_{max}=\sqrt{2I_{laser}/(c^2\epsilon_0)}}$, $\mathrm{k_L=2\pi/\lambda_L}$, and $\mathrm{\omega_L=ck_L}$ with $\mathrm{\lambda_L=527 \text{nm}}$. The analytic solution for the electric and magnetic fields for such a counter-propagating laser is shown below. A snapshot of the resulting standing wave is shown in Fig. \ref{fig:ezby_snapshot}.  

\begin{eqnarray}
\mathrm{\mathbf{E}_L}=\mathrm{2E_0sin(k_Lx)sin(\omega_L t)}\mathbf{\hat{z}} \nonumber \\
\mathrm{\mathbf{B}_L}=\mathrm{2B_0cos(k_Lx)cos(\omega_L t)}\mathbf{\hat{y}}.
\label{eq:em_standing}
\end{eqnarray}

The fluid-approximation validity is maintained by ensuring the Langdon parameter is kept low enough ($\alpha<1$) to avoid electrons becoming super-Gaussian due to the well-known Langdon effect \cite{langdon_nonlinear_1980, turnbull_impact_2020}.   
 
When $\mathbf{j}$ is computed using Eq. (\ref{eq:jtoE}) and simplifying assumptions of Eq. (\ref{eq:GOLPhasorDC}), the current in Eq. (\ref{eq:dyadic_phasor1}) can be written as,
\begin{equation}
    \mathrm{\mathbf{j}=-\frac{2E_0\epsilon_0\omega_{pe}^2}{\omega_L}sin(kx)cos(\omega_Lt)\mathbf{\hat{z}}}. 
    \label{eq:fast_j}
\end{equation}

The resulting Hall term in the single-fluid GOL is now easily computed using Eq. (\ref{eq:fast_j}) and the analytic solution for the standing wave for both $\mathbf{j}$ and $\mathbf{B}$.  As this is a source term for ion-fluid momentum conservation in Eq. (\ref{eq:momentum_eq}),  it really represents the effect on an ion-fluid from time-averaged electron dynamics.

 The ponderomotive force is only related to the Hall term averaged over time ($\langle\mathbf{j}\times\mathbf{B}\rangle$), which will result in a factor of $\frac{1}{2}$. 
 With the assumption that $\mathrm{\Vert{\mathbf{B}}\Vert\approx\Vert{\mathbf{E}}\Vert/c}$ and a trigonometric identity, the time-averaged ponderomotive term of the GOL is simplified further. 

 \begin{eqnarray}
\mathrm{\mathbf{F}_{\langle Hall\rangle}}=\mathrm{-\frac{E_0^2e^2N_e\lambda_L}{2\pi m_ec^2}sin(2kx)} \mathbf{\hat{x}}\nonumber \\
\label{eq:hall_term_1}
\end{eqnarray}

\begin{figure}[ht]
\centerline{\includegraphics[]{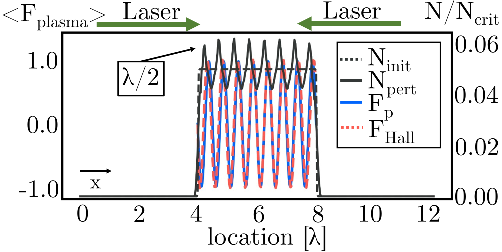}}
\caption{\label{fig:ponderomotive_steepening_theory_resist}Shown is a representative profile ion density modulation caused by ponderomotive force on a plasma obeying a simple Spitzer resistivity model.  The dotted black line indicates the initial density profile, while the solid line represents the same profile but much later in time ($\mathrm{t\approx675\tau_{L}}$).   The blue solid line represents the theoretical ponderomotive force and the dotted line represents the time-averaged Hall force ($\mathrm{\mathbf{j}\times\mathbf{B}}$).}
\end{figure}

Using the simulation parameters described above, the windowed cycle-averaged $\mathrm{\mathbf{j\times B}}$ was computed and plotted in Fig. \ref{fig:ponderomotive_steepening_theory_resist} along with the theoretical cycled-averaged ponderomotive force (computed for the electron-fluid).

The initial plasma density profile is also plotted along with one from much later in the simulation ($\mathrm{t\approx675\tau_{L}}$).  The results of the theory and simulation of the ponderomotive force agree very well, and there is density steepening and a comb-like appearance described in earlier referenced works\cite{lehmann_plasma_2019, sheng_plasma_2003, plaja_analytical_1997}.  It should also be noted that in a prior work by these authors \cite{young_impact_2024}, the same steepening was observed when comparing a PIC model (OSIRIS\cite{hemker_particle_cell_2015}) to our XMHD model, and they also matched very closely. 

The steepening is significantly enhanced when resistivity is present.  This is likely due to easier thermalization of the directed kinetic energy attained by the electrons as they wiggle in the laser fields. This energy then appears as a gradient in $T_e$, and therefore a $\nabla\cdot \mathrm{p_e}\mathbf{I}$, which would further push ions away from regions of high fields. 

\begin{figure}[ht]
\centerline{\includegraphics[]{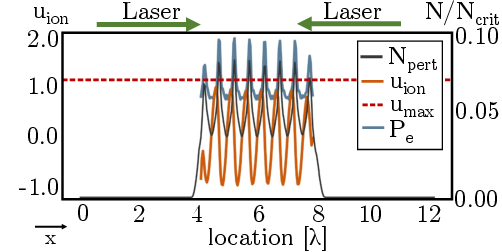}}
\caption{\label{fig:ponderomotive_ionvel} Shown are the ion velocity $\mathrm{u_{ion}}$, an analytic estimate of maximum ion velocity $\mathrm{u_{max}}$ late in time ($\mathrm{t\approx1420\tau_{L}}$), ion density modulation $\mathrm{N_{pert}}$, and electron pressure $\mathrm{P_e}$ caused by ponderomotive force on plasma with Spitzer resistivity. }
\end{figure}

The ion velocity is the final piece of evidence to demonstrate XMHD captures the physics in Eq. (\ref{eq:velocity_int}).  Fig. \ref{fig:ponderomotive_ionvel} shows ion-velocity in the direction of ponderomotive force ($\mathrm{u_x}$) along with $\mathrm{u_{max}}$ from Eq. (\ref{eq:velocity_int}).  The magnitude is similar and the $\mathrm{u_x}$ peak matches the previous analytic theory, which suggested $\mathrm{\lambda_L/8}$.  The ion pressure and temperature also peak at $\mathrm{x=\lambda_L/4}$ as suggested earlier.  This implies there will be a counterbalance to ponderomotive force at the ion peaks since $\mathrm{\nabla\mathrm{p}}\propto\langle \mathbf{j}\times \mathbf{B}\rangle$.  As the ponderomotive force acts on the ion fluid, the pressure continues to build and oppose a further increase in velocity. 

\begin{figure}[ht]
\centerline{\includegraphics[]{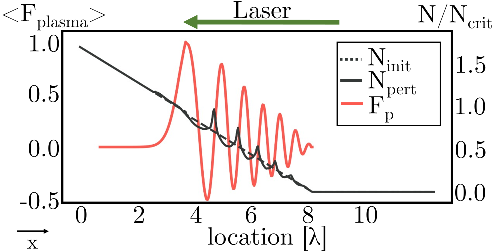}}
\caption{\label{fig:ponderomotive_steepening_ramp} The beginning of an ion density modulation caused by ponderomotive force on a ramped plasma.  The dotted black line indicates the initial density profile, while the solid line represents the same profile but much later in time ($\mathrm{t\approx577\tau_{L}}$).  }
\end{figure}

The results in Fig. \ref{fig:ponderomotive_steepening_ramp} use a single laser, sent from the +x-axis to reflect off a steep density ramp (similar to the setup in an earlier PIC work \cite{smith_particle--cell_2019}).  The reflection causes its own wave-beating and ensuing ponderomotive effects such as density steepening. 

This figure also shows that ponderomotive effects qualitatively increase with density, as previously suggested.  Much later in time, this effect causes supercritical steepening and plateauing of the density profile, which renders the interior of the plasma inaccessible to the laser while also potentially providing new supercritical surfaces for increased EM-reflection \cite{estabrook_two-dimensional_1975}. 

In addition to a forward moving supercritical shock, there are possibly 1 or more supercritical density perturbations even closer to the laser source (troughs in Fig. \ref{fig:ponderomotive_steepening_ramp}).  The cavity between these can also capture EM radiation and increase absorption.

To verify this is a valid solution for the ions, we compare this to the ponderomotive force from theory, the dimensional form of Eq. \ref{eq:Fpond}.  When a standing-wave electric field solution of Eq. \ref{eq:em_standing} is substituted into this, a closed-formed solution for $\mathrm{\mathbf{F}_p}$ is attained.   


\begin{flalign}
    \mathrm{\mathbf{F}_p} & \mathrm{=-\frac{e^2}{4m\omega^2}\nabla(E^2)} \nonumber \\
     &=-\mathrm{\frac{e^2E_0^2\lambda_L}{2\pi mc^2}sin(2k_Lx)} \mathbf{\hat{x}}\nonumber \\
    \label{eq:ponderomotive2}
\end{flalign}


The result in Eq. (\ref{eq:ponderomotive2}) matches Eq. (\ref{eq:hall_term_1}), except for $m$ in the denominator and the absence of $\mathrm{N_e}$ in the numerator. The ponderomotive force is not directly applied to ions with $\mathrm{m=m_{i}}$, as this force is negligible compared to $\mathrm{m=m_e}$. Single fluid XMHD model avoids the intermediate step of electron density displacement for the ponderomotive force to act on an ion. It assumes electrons move until the Coulomb force equals the ponderomotive force on the electrons, equating the force on ion-fluid to electron-fluid and replacing $\mathrm{m=m_e}$. The $\mathrm{N_e}$ in the numerator indicates it's a force on the collective electron population. Our result agrees with the theory and PIC simulation in Ref. \cite{smith_particle--cell_2019}, which uses a similar setup.


\section{Conclusion}

This study demonstrates native ponderomotive effects in a single-fluid, two-energy extended magnetohydrodynamics (XMHD) model by averaging electron dynamics over an ion timescale. The model captures key phenomena without complex multi-fluid or particle-based simulations. Theoretical analysis and one-dimensional simulations support the model’s validity, aligning with analytical theories and particle-in-cell (PIC) simulations. 

The model replicates critical ponderomotive effects like density modulation, which leads to laser self-focusing, stimulated Brillouin scattering (SBS), and crossed-beam energy transfer (CBET). 

Future work could explore incorporating charge density fluctuations to enable electron plasma waves and resonant absorption, enhancing the model's scope. Since prior work shows that full knowledge of plasma conditions leads to accurate simulations \cite{turnbull_refractive_2017,turnbull_impact_2020}, then collisional models, easily tested with fluid codes \cite{young_impact_2024}, suggest resistivity as a future parameter to explore.

This model impacts next-generation laser systems and fusion research by providing a more accessible and efficient means to simulate complex plasma such as combined ponderomotive and collisional absorption effects.

\section{Funding}
This research was supported by the NSF CAREER Award PHY-1943939, the DOE center DE-NA0004148, and by the Laboratory for Laser Energetics Horton Fellowship.

\bibliography{bibliography}
\end{document}